\documentclass{article}
\pdfoutput=1
 \usepackage{siunitx} 

 \usepackage[dvips]{graphicx} 
\usepackage{latexsym}
\usepackage{amsmath}
\usepackage{amsthm}
\usepackage{mathrsfs}
\usepackage[numbers]{natbib}
\let \chapter \section
\usepackage[ruled, vlined, linesnumbered]{algorithm2e}
\usepackage[font=bf, labelfont=bf]{caption}
\usepackage{enumerate}
\usepackage{enumitem}
\usepackage{multirow}
\usepackage{color}
\usepackage{xfrac}
\usepackage{lmodern}

\usepackage{amssymb}
\usepackage{booktabs}
\usepackage{mathrsfs}
\usepackage{enumerate}
\usepackage{color}
\usepackage{enumitem}
\usepackage{pbox}
\usepackage{rotating}
\usepackage{url}
\usepackage{authblk}

\graphicspath{{./figs/}}

\newcommand{\argmax}{\operatornamewithlimits{argmax}}

\newcommand{\attr}{\mathit{attr}}
\newtheorem{definition}{Definition}

\title{Generating Realistic Synthetic Population Datasets}

\author{Hao Wu\textsuperscript{1,3},
Yue Ning\textsuperscript{2,3},
Prithwish Chakraborty\textsuperscript{2,3}\\
Jilles Vreeken\textsuperscript{4,5},
Nikolaj Tatti\textsuperscript{6},
Naren Ramakrishnan\textsuperscript{2,3}
}
\affil{\textsuperscript{1}Department of Electrical and Computer Engineering,
Virginia Tech, USA \\
\textsuperscript{2}Department of Computer Science, Virginia Tech, USA \\
\textsuperscript{3}Discovery Analytics Center, Virginia Tech, USA \\
\textsuperscript{4}Max Planck Institute for Informatics, Saarbr{\"u}cken,
Germany \\
\textsuperscript{5}Cluster of Excellence MMCI, Saarland University,
Saarbr{\"u}cken, Germany \\
\textsuperscript{6}HIIT, Department of Information and Computer Science, Aalto University, Finland}

\date{}

\begin{document}

\maketitle

\begin{abstract}
	Modern studies of societal phenomena rely on the availability of large datasets
capturing attributes and activities of synthetic, city-level, populations.  For
instance, in epidemiology, synthetic population datasets are necessary to study
disease propagation and intervention measures before implementation. In social
science, synthetic population datasets are needed to understand how policy
decisions might affect preferences and behaviors of individuals. In public
health, synthetic population datasets are necessary to capture diagnostic and
procedural characteristics of patient records without violating
confidentialities of individuals.  To generate such datasets over a large set of
categorical variables, we propose the use of the maximum entropy principle to
formalize a generative model such that in a statistically well-founded way we
can optimally utilize given prior information about the data, and are unbiased
otherwise. An efficient inference algorithm is designed to estimate the maximum
entropy model, and we demonstrate how our approach is adept at estimating
underlying data distributions. We evaluate this approach against both simulated
data and on US census datasets, and demonstrate its feasibility using an
epidemic simulation application.

\end{abstract}

\section{introduction}
\label{sec:intro}
Many research areas, e.g., epidemiology, public health, social science, study
the behavior of large populations of individuals under natural scenarios as well
as under human interventions. A key need across these domains is the ready availability 
of realistic synthetic datasets that can capture key attributes and activities of large populations.

For instance, in epidemiology, synthetic populations are necessary to
study disease propagation and intervention measures
before implementation. Information from the US census is typically used to model
such synthetic datasets.  In social science, synthetic populations are necessary to
understand how policy decisions might affect preferences and behaviors of
individuals. Finally, in public health, synthetic populations are necessary to
capture diagnostic and procedural characteristics of patient records without
violating confidentialities of individuals. 

Typically, the constraints underlying synthetic population generation are assumptions
on the supporting marginal or conditional distributions.  Although there exist
prior research in estimating probability distributions subject to constraints
(e.g., Monte Carlo methods), they are primarily focused on continuous-valued
data. Many domains on the other hand, such as those studied here, feature the
need for multi-dimensional categorical datasets.

As a case in point, in epidemiology, one important task is to simulate disease
spread and potential outbreaks on the city- or nation-level, and provide useful
information to public health officials to support policy and decision making. To
make such simulations as accurate as possible, synthetic populations that have
the same structural and behavioral properties as the real population are needed.
In domains like health care, privacy is an additional issue motivating the
design of synthetic populations.  In these applications, the necessary datasets
to be generated can be represented as tuples with categorical data attributes.

Motivated by these emerging needs, we focus our attention on constructing a
generative model that captures given characteristics of categorical population attributes, and
best estimates the underlying data generation distribution. However, modeling
multi-dimensional categorical data and estimating distributions can be quite
challenging due to the exponential possibilities of data spaces in terms of the
number of dimensions of categorical data tuples. To address these challenges
and difficulties, we take the first step here to study this problem. To model
categorical data with statistical constraints, we apply the classical and
statistically well-founded maximum entropy model. We construct a generative
maximum entropy model wherein the probabilities of certain categorical patterns
are required to satisfy given constraints. In this way, the maximum entropy
model maintains the selected characteristics of the underlying categorical data
distribution. By sampling the categorical tuples from the maximum entropy model,
synthetic population datasets can be generated.

\begin{figure}
	\centering
	\includegraphics[width=2.5in]{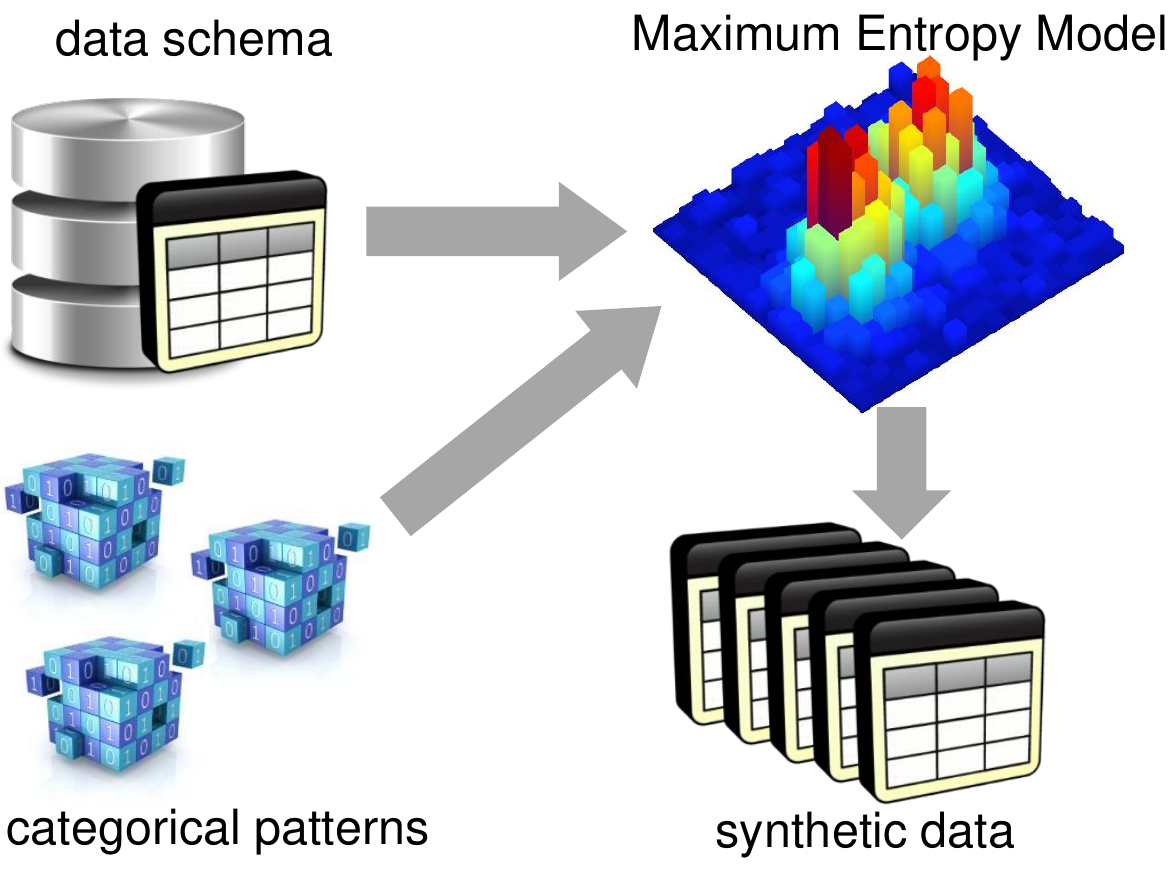}
	\caption{Process of generating realistic synthetic data with our proposed
		approach.}
	\label{fig:1.1}\vspace{-0.2in}
\end{figure}

Generally, solving maximum entropy models can be infeasible in practice. In this
paper, we show that by leveraging the structure of the categorical data space in
our setting, the maximum entropy model could be inferred quite efficiently. We
also propose a heuristic together with the Bayesian information criterion (BIC)
to select a simple as well as an informative model. To summarize our approach in
a nutshell, our contributions are:
\begin{enumerate}[itemsep=-2pt, topsep=-2pt]
	\item We formalize the problem of generating synthetic population datasets
		via a generative maximum entropy model for categorical data, which
		captures the statistical features of the underlying categorical data
		distributions.
	\item By exploring the structure of the categorical data space, we propose a
		partition scheme to make the maximum entropy model inference more
		efficient than the general case. We also present an efficient
		graph-based model inference algorithm.
	\item We propose a BIC-based heuristic to perform model selection wherein
		the simple and informative maximum entropy model will be chosen.
	\item Using results on both synthetic datasets and real US census data, we
		demonstrate that the proposed maximum entropy model is capable of
		recovering the underlying categorical data distribution and generating
		relevant synthetic populations.
\end{enumerate}

\section{Preliminaries}
\label{sec:pre}
Let $\mathcal{A} = \{A_1, A_2, \ldots, A_q\}$ denote a set of categorical random
variables (or attributes), and $\mathcal{R}(A_i) = \{a_1^{(i)}, a_2^{(i)},
\ldots, a_{k_i}^{(i)}\}$ represent the set of $k_i$ possible values for random
variable $A_i$. Here, $|\cdot|$, e.g.\ $|\mathcal{R}(A_i)|$, is used to
represent the cardinality of a set.

By a random categorical tuple, we mean a vector of categorical random
variables, e.g. $T = (A_1, A_2, \ldots, A_q)$, which is generated by some
unknown probability distribution. The notation of $T(A_i)$ is used to represent
the value of attribute $A_i$ in tuple $T$. The space of all the possible
categorical tuples is denoted by $\mathcal{S} = \prod_{i=1}^{q}
\mathcal{R}(A_i)$, where $\prod \cdot$ is the series of Cartesian product over
the given sets.  Given a categorical pattern, which is defined as an ordered set
$X = (A_i \mid A_i \in C, C \subseteq \mathcal{A})$ over a subset of random
variables $C \subseteq \mathcal{A}$, let $\mathcal{S}_{X} = \prod_{A_i \in C}
\mathcal{R}(A_i)$ represent the space that contains all the possible values of
pattern $X$. An instantiation of pattern $X$ is defined as $\boldsymbol{x} =
\left(a_j^{(i)} \mid a_j^{(i)} \in \mathcal{R}(A_i), A_i \in C, C \subseteq
\mathcal{A}\right)$, and $X(A_i)$ is used to represent the value of attribute
$A_i$ in the pattern $X$.

For any pattern value $\boldsymbol{x}$ associated with pattern $X$, we use the
notation of $T = \boldsymbol{x}$ if the corresponding random variables in $T$
equal to the values in $\boldsymbol{x}$ and $p(T = \boldsymbol{x})$ to denote
the probability of $T = \boldsymbol{x}$. Given a categorical dataset $D$,
$\tilde{p}(T = \boldsymbol{x} \mid D)$ is used to denote the empirical
probability of $T = \boldsymbol{x}$ in the dataset $D$. An indicator function
$I_X(T=\boldsymbol{x}) : \mathcal{S} \rightarrow \{0, 1\}$ of pattern $X$, which
maps a categorical tuple to a binary value, is defined as:
$$
I_X(T = \boldsymbol{x}) = \left\{ 
	\begin{array}{l}
		1, \quad \text{if}~T = \boldsymbol{x}, \\
		0, \quad \text{otherwise}.
	\end{array} \right.
$$

Given a probability distribution $p$ over the categorical tuple space
$\mathcal{S}$, the entropy $H (p)$ with respect to $p$ is defined as:
$$
H (p) = -\sum_{T \in{} \mathcal{S}} p (T) \log{p (T)}~.
$$
The Maximum Entropy principle states that among a set of probability
distributions $\mathcal{P}$ that comply with the given prior information about
the data, the maximum entropy distribution
$$
p^{*} = \argmax_{p \in{} \mathcal{P}} H (p)
$$
will optimally use the current prior information and best summarize the data. 
Otherwise, it is fully unbiased.

\paragraph{Problem Statement}
Given a set of categorical patterns $\mathcal{X}$ with associated empirical
frequencies as the prior information of a dataset, we would like to find a
probabilistic model $p$ that best utilizes such prior information and helps to
regenerate categorical datasets that conform to the given prior information.

\section{Categorical Maximum Entropy model}
\label{sec:cate_maxent}

\subsection{Categorical MaxEnt Model Specification}
\label{sec:model_spec}

Suppose we have a set categorical patterns $\mathcal{X} = \{X_i \mid i = 1, 2,
\ldots, n\}$ and an associated set of empirical probabilities $\tilde{P} =
\{\tilde{p}(T = \boldsymbol{x}_{i,j} \mid D) \mid \boldsymbol{x}_{i,j} \in{}
\mathcal{S}_{X_i}, i = 1, 2, \ldots, n\}$ as prior information about dataset
$D$. Here, $\boldsymbol{x}_{i,j}$ denotes the $j^{th}$ value of the pattern
$X_i$. Notice that it is not necessary that every possible value of
pattern $X_i$ in $\mathcal{S}_{X_i}$ is provided as part of the prior
information here. Such prior information identifies a group of probability
distributions $\mathcal{P}$ over $\mathcal{S}$ which agree with the empirical
probabilities of the given categorical patterns. That is:
\begin{align}
	\label{eq:3.A.0}
	\mathcal{P} & = \{ p \}~\text{s.t.}~p(T = \boldsymbol{x}_{i,j}) =
	\tilde{p}(T = \boldsymbol{x}_{i,j} \mid D), \\
	& \forall p \in \mathcal{P}, X_i \in \mathcal{X},~\text{and}~\tilde{p}(T =
	\boldsymbol{x}_{i,j} \mid D) \in \tilde{P} \nonumber
\end{align}
Following the Maximum Entropy principle, for all $p \in{} \mathcal{P}$, we are
particularly interested in the Maximum Entropy distribution which optimally
represents the given prior information. The famous theorem
in~\cite{csiszar:75:i-divergence} (Theorem 3.1) shows that the Maximum Entropy
distribution has an exponential form. In our categorical scenario, the Maximum
Entropy distribution could be written as
\begin{align}
	p^{*}(T) = u_0 \prod_{X_i \in{} \mathcal{X}} \prod_{\boldsymbol{x}_{i,j}
	\in \mathcal{S}_{X_i}}
	{\left(u_{i,j}\right)}^{I_{X_i}(T=\boldsymbol{x}_{i,j})}~,
\label{eq:3.A.1}
\end{align}
where $u_{i,j} \in \mathbb{R}$ are the model parameters associated with each
model constraint specified in Equation~\eqref{eq:3.A.0}, and $u_0$ is the
normalizing constant.

\subsection{Incorporating Individual Attribute Frequ-encies}
\label{sec:column_margin}
The frequencies of individual attributes play an important role in the pattern
analysis and discovery. Such frequencies characterize the attribute marginal
distributions which convey basic information about the data currently under
investigation, and yet are relatively easy to calculate from the data.
Incorporating such individual attribute frequencies will enrich the categorical
Maximum Entropy model, and make it more informative.

Although such individual attribute frequencies can be trea-ted as part of the
categorical pattern set $\mathcal{X}$, considering the computation efficiency
which will be explained in detail in the next section, the categorical Maximum
Entropy model treats them separately. Let $v_{i,j}$ denote the model parameters
corresponding to the individual attribute model constraints, then, the Maximum
Entropy distribution can be factorized as:
\begin{align}
	\label{eq:3.B.1}
	p^{*}(T) = u_0 \prod_{X_i \in{} \mathcal{X}} \prod_{\boldsymbol{x}_{i,j}
	\in \mathcal{S}_{X_i}}
	{\left(u_{i,j}\right)}^{I_{X_i}(T=\boldsymbol{x}_{i,j})} \times \prod_{A_i
		\in{} \mathcal{A}} \prod_{a_j \in{} \mathcal{R}(A_i)}
		{\left(v_{i,j}\right)}^{I_{A_i}(T=a_j)}~.
\end{align}
Notice that the second component involved with $v_{i,j}$ also follows the
exponential form described in Equation~\eqref{eq:3.A.1}. By introducing a
normalizing constant $v_0$, an independent Maximum Entropy distribution
$p_{\mathcal{A}}(T)$ that only involves individual attribute constraints could
be defined as:
\begin{align}
	p_{\mathcal{A}}(T) = v_0 \prod_{A_i \in{} \mathcal{A}} \prod_{a_j \in{}
	\mathcal{R}(A_i)} {\left(v_{i,j}\right)}^{I_{A_i}(T=a_j)}~.
	\label{eq:3.B.2}
\end{align}
Combining Equation~\eqref{eq:3.B.1} and~\eqref{eq:3.B.2}, the Maximum Entropy
distribution that incorporates individual attribute frequencies would be
specified as:
\begin{align}
	p^{*}(T) = p_{\mathcal{A}}(T) \frac{u_0}{v_0} \prod_{X_i \in{} \mathcal{X}}
	& \prod_{\boldsymbol{x}_{i,j} \in \mathcal{S}_{X_i}}
	{\left(u_{i,j}\right)}^{I_{X_i}(T=\boldsymbol{x}_{i,j})}~.
	\label{eq:3.B.3}
\end{align}

\section{Model Inference}
\label{sec:infer}
In this section, we develop an efficient algorithm to infer the categorical
Maximum Entropy model. Our algorithm is built on the well-known Iterative
Scaling~\cite{1972:iterative:scaling} framework. The general idea of the
algorithm is that starting from the uniform distribution, it iteratively updates
each model parameter to make the distribution satisfy the corresponding
constraint until it converges to the Maximum Entropy distribution.

\subsection{Efficient Model Inference}
\label{sec:model_infer}
The main challenge in the Iterative Scaling framework is how to efficiently
query the Maximum Entropy model during the iterative updates of the model
parameters. In order to achieve that, we need to explore the particular
structure of the tuple space $\mathcal{S}$ determined by the given pattern set
$\mathcal{X}$. After examining the exponential form of the Maximum Entropy
distribution in Equation~\eqref{eq:3.A.1}, we observe that for any two
categorical tuples $T_1$ and $T_2$ in $\mathcal{S}$, if they contain the
same subset of categorical patterns in $\mathcal{X}$, they will have the same
probability under the Maximum Entropy distribution inferred $\mathcal{X}$. In
another word, $\forall T_1, T_2 \in \mathcal{S}$, if $I_{X_i}(T_1 =
\boldsymbol{x}_{i,j}) = I_{X_i}(T_2 = \boldsymbol{x}_{i,j})$ holds true for all
$X_i \in \mathcal{X}$ and $\tilde{p}(T = \boldsymbol{x}_{i,j} \mid D) \in
\tilde{P}$, then $p^*(T_1) = p^{*}(T_2)$. Based on such observation, we have
the following definition of tuple block.
\begin{definition}
	A tuple block $B$ is a set categorical tuples such that $\forall T_1, T_2
	\in B$, $I_{X_i}(T_1 = \boldsymbol{x}_{i,j}) = I_{X_i}(T_2 =
	\boldsymbol{x}_{i,j})$ holds true for all $X_i \in \mathcal{X}$,
	$\boldsymbol{x}_{i,j} \in \mathcal{S}_{X_i}, and~
	\tilde{p}(T=\boldsymbol{x}_{i,j} \mid D) \in \tilde{P}$.
\end{definition}

With the definition of tuple block, we could partition the entire
categorical tuple space into several tuple blocks. When
$|\mathcal{X}| \ll |\mathcal{A}|$, the partition scheme introduced here could
greatly reduce the dimensionality of the space we are working on. Here, we use
$\mathcal{B}_{\mathcal{X}}$ to denote the tuple block space generated
based on pattern set $\mathcal{X}$. Also, the definition of tuple block
let us extend the indicator function defined over tuple space to the
domain of tuple block, which is defined as:
$$
I_{X_i}(B \mid \boldsymbol{x}_{i,j}) = I_{X_i}(T =
\boldsymbol{x}_{i,j}), \quad \forall X_i \in \mathcal{X}, T \in B.
$$

\begin{algorithm}[t]
	\SetAlgoLined
	\SetKwInOut{Input}{input}
	\SetKwInOut{Output}{output}
	\SetKwFunction{KwFnFind}{findPosition}
	\SetKwFunction{KwFnMerge}{createBlock}

	\Input{A set of categorical patterns $\mathcal{X}$, and associated empirical
	probabilities $\tilde{P}$.}
	\Output{tuple block graph $G$.}
	\BlankLine

	Let $G \leftarrow \{\varnothing\}$\;\label{graph:line:1}
	\ForEach{$X_i \in \mathcal{X}, \boldsymbol{x}_{i,j} \in \mathcal{S}_{X_i}
	~s.t.~\tilde{p}(T = \boldsymbol{x}_{i,j}) \in \tilde{P}$}{
		\ForEach{$B_k \in G$}{
			$B_{\mathit{new}} \leftarrow$ \KwFnMerge{$B_k,
			X_i$}\;\label{graph:line:4}
			\If{$B_{\mathit{new}} \neq \mathit{Null}$}{
				\KwFnFind{$\varnothing$, Null,
					$B_{\mathit{new}}$}\;\label{graph:line:6} }
		}
	}
	\Return{$G$}\;
	\caption{Constructing tuple Block Graph}
	\label{alg:4.A.2}
\end{algorithm}

\begin{algorithm}[t]
	\SetAlgoLined
	\SetKwInOut{Input}{input}
	\SetKwInOut{Output}{output}
	\SetKwFunction{KwFnFind}{findPosition}
	\SetKwFunction{KwFnCheck}{checkDescendant}
	\SetKwFunction{KwFnInsert}{InsertDescendant}
	\SetKwProg{KwSubRoutine}{Procedure}{:}{}

	\Input{Current block $B_{\mathit{curr}}$, last visited block
	$B_{\mathit{last}}$, new block $B_{\mathit{new}}$.}
	\Output{Success or Fail.}
	\BlankLine
	
	\uIf{$B_{\mathit{new}}$ and $B_{\mathit{curr}}$ are the
		same}{\label{find:line:26}
		\Return{Success}\;\label{find:line:27}
	}
	\uElseIf{$B_{\mathit{new}} \subseteq
	B_{\mathit{\mathit{curr}}}$}{\label{find:line:21}
		$\mathit{child}(B_{\mathit{last}}) \leftarrow
		\mathit{child}(B_{\mathit{last}}) \setminus \{B_{\mathit{curr}}\}$\;
		$\mathit{child}(B_{\mathit{new}}) \leftarrow
		\mathit{child}(B_{\mathit{new}}) \cup \{B_{\mathit{curr}}\}$\;
		$\mathit{child}(B_{\mathit{last}}) \leftarrow
		\mathit{child}(B_{\mathit{last}}) \cup \{B_{\mathit{new}}\}$\;
		\Return{Success}\;\label{find:line:25}
	}
	\uElseIf{$B_{\mathit{curr}} \subseteq B_{\mathit{new}}$}{
		\uIf{$\mathit{child}(B_{\mathit{curr}}) =
			\emptyset$}{\label{find:line:2} 
			$\mathit{child}(B_{\mathit{curr}}) \leftarrow
			\mathit{child}(B_{\mathit{curr}}) \cup \{B_{\mathit{new}}\}$\;
			\Return{Success}\;\label{find:line:4}
		}\uElse{
			$\mathit{failBlock} \leftarrow$ \KwFnInsert{$B_{\mathit{new}}$,
			$B_{\mathit{curr}}$}\;
			\KwFnCheck{$\mathit{failBlock}$,
			$B_{\mathit{new}}$}\;\label{find:line:18}
			\Return{Success}\;
		}
	}
	\Return{Fail}\;\label{find:line:29}
	\BlankLine
	\KwSubRoutine{\KwFnInsert{$B_{\mathit{new}}$, $B_{\mathit{curr}}$}}{
		$\mathit{failBlock} \leftarrow \emptyset$, $\mathit{accu} \leftarrow
		$ Fail\;
		\ForEach{$B_k \in \mathit{child}(B_{\mathit{curr}})$}{\label{find:line:6}
			$r \leftarrow $ \KwFnFind{$B_k$, $B_{\mathit{curr}}$,
			$B_{\mathit{new}}$}\; 
			\uIf{$r = $ Success}{
				$\mathit{accu} \leftarrow $ Success\;
			}\uElse{
				$\mathit{failBlock} \leftarrow \mathit{failBlock} \cup
				\{B_k\}$\;\label{find:line:14}
			}
		}
		\uIf{$\mathit{accu} = $ Fail}{\label{find:line:15}
			$\mathit{child}(B_{\mathit{curr}}) \leftarrow
			\mathit{child}(B_{\mathit{curr}}) \cup
			\{B_{\mathit{new}}\}$\;\label{find:line:17}
		}
		\Return{$\mathit{failBlock}$}\;
	}

	\caption{\textit{findPosition} procedure}
	\label{alg:4.A.3}
\end{algorithm}

By introducing tuple blocks, we transfer the problem of computing
categorical pattern probability $p(T = \boldsymbol{x}_{i,j})$ on tuple
space to the block space, which makes it possible to calculate $p(T =
\boldsymbol{x}_{i,j})$ in a reasonable time. In the context of tuple
blocks, the pattern probability $p(T = \boldsymbol{x}_{i,j})$ in would be
$$
p(T = \boldsymbol{x}_{i,j}) = \sum_{\substack{B \in \mathcal{B}_{\mathcal{X}},
\\ I_{X_i}(B \mid \boldsymbol{x}_{i,j}) = 1}} p(B)~,
$$
where $p(B)$ is the probability for tuple block $B$. Since the
probabilities for the categorical tuples within the same block are all the
same, the probability for the tuple block $B$ is defined as: 
$$
p(B) = \sum_{T \in B} p(T) = |B| \times u_0 \prod_{X_i \in \mathcal{X}}
\prod_{\boldsymbol{x}_{i,j} \in \mathcal{S}_{X_i}} (u_{i,j})^{I_{X_i}(B \mid
\boldsymbol{x}_{i,j})}~.
$$

Now, our problem comes down to how to organize the tuple block space
$\mathcal{B}_{\mathcal{X}}$ and efficiently compute the number of
categorical tuples in each block, or in other words, the size $|B|$ of
each tuple block $B$. In order to achieve that, we introduce a partial order
on $\mathcal{B}_{\mathcal{X}}$. Let
$$
\attr(B) = \bigcup_{\substack{X_i \in \mathcal{X}, \\ I_{X_i}(B \mid
\boldsymbol{x}_{i,j}) = 1}} X_i~,
$$
which represents the set of attributes involved by tuple block $B$. Then,
we have the definition about the partial order over $\mathcal{B}_{\mathcal{X}}$
as described below.
\begin{definition}
	\label{def:2}
	Given any tuple blocks $B_1, B_2 \in \mathcal{B}_{\mathcal{X}}$,
	$B_1 \subseteq B_2$ if and only if the following conditions hold true:
	\begin{enumerate}[itemsep=-2pt, topsep=-2pt]
		\item $\attr(B_1) \subseteq \attr(B_2)$;
		\item $B_1(A_k) = B_2(A_k),~\forall A_k \in \attr(B_1) \cap \attr(B_2)$.
	\end{enumerate}
\end{definition}
\noindent Here, $B(A_k)$ denotes the value of attribute $A_k$ in the tuple
block $B$. It is easy to verify that Definition~\ref{def:2} satisfies the
property of reflexivity, antisymmetry and transitivity.

With the partial order $\subseteq$ defined on $\mathcal{B}_{\mathcal{X}}$ here,
it is natural to organize the tuple blocks into a hierarchical graph structure.
That is, if tuple block $B_k \subseteq B_l$, block $B_l$ is organized as the
child of block $B_k$. Algorithm~\ref{alg:4.A.2} illustrates how such block graph
is constructed and maintained. The algorithm starts with the graph that has only
one block represented by $\varnothing$ indicating that none of the categorical
patterns is involved in this block (line~\ref{graph:line:1}). We will refer this
block as root block in the rest of this section. Then, for each of the pattern
set $X_i \in \mathcal{X}$ and its possible value $\boldsymbol{x}_{i,j}$, we
attempt to create a new tuple block by merging it with every existing block
$B_k$ from root level to leaf level (without child blocks) in the current block
graph $G$ if they are compatible (line~\ref{graph:line:4}). A categorical
pattern $X_i$ is not compatible with tuple block $B_k$ if $\attr(B_k) \cap X_i
\neq \emptyset$, and $\exists A_i \in \attr(B_k) \cap X_i$ such that $B_k(A_i)
\neq X_i(A_i)$. If a new tuple block $B_{\mathit{new}}$ is created, it is
obvious that for all $X_l \in \mathcal{X}, I_{X_l}(B_k \mid
\boldsymbol{x}_{l,j}) = 1$, we have $I_{X_l}(B_{\mathit{new}} \mid
\boldsymbol{x}_{l,j}) = 1$ and also $I_{X_i}(B_{\mathit{new}} \mid
\boldsymbol{x}_{i,j}) = 1$. Finally, the new tuple block $B_{\mathit{new}}$ will
be added into the current block graph $G$ based on the partial order described
in Definition~\ref{def:2} (line~\ref{graph:line:6}).

To be more specific, Algorithm~\ref{alg:4.A.3} illustrates how the procedure
\textit{findPosition} inserts a new tuple block into the block graph $G$
in a recursive manner. Depending on the relationship between the current block
$B_{\mathit{curr}}$ we are visiting and the new block $B_{\mathit{new}}$, the
insertion operation could be classified into four scenarios.\\

\begin{figure*}[t]
	\begin{minipage}[t]{1.0\textwidth}
		\centering
		\includegraphics[width=1.0\textwidth]{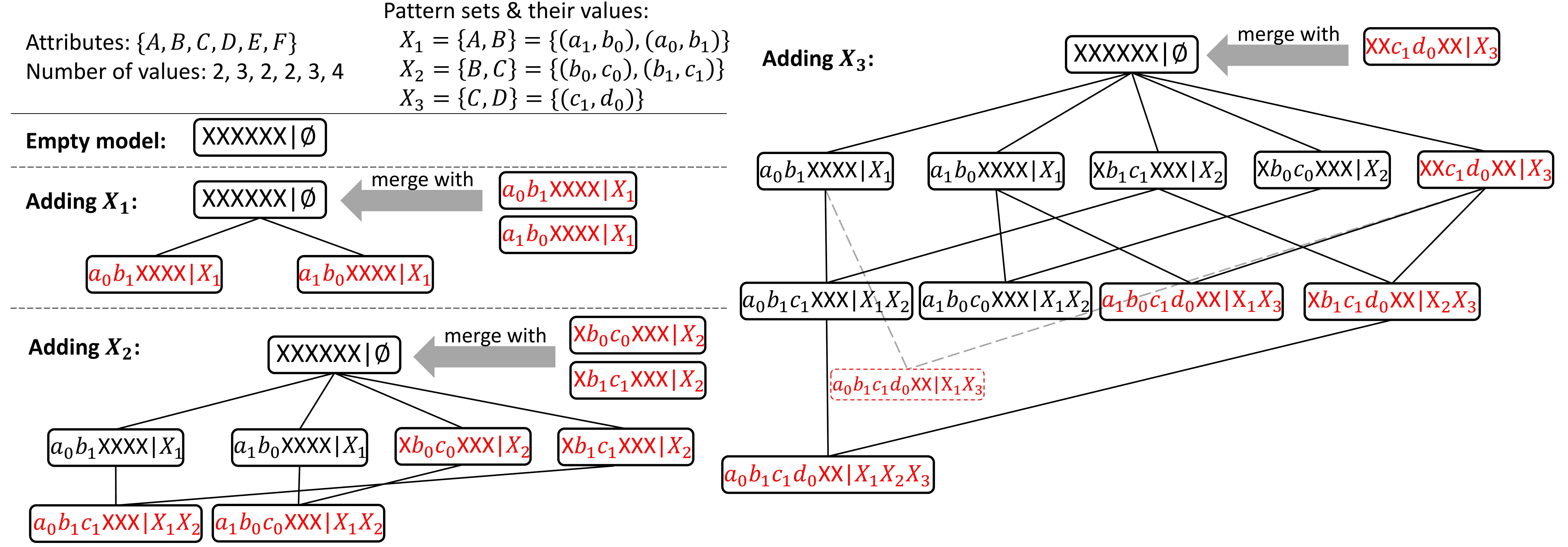}
		\caption{Example of constructing tuple block graph on
		toy dataset with 6 attributes and 3 categorical patterns. The blocks
		marked with red denote the new tuple blocks created in each
		iteration by adding new categorical patterns.}
		\label{fig:4.A.1}\vspace{-0.2in}
	\end{minipage}
\end{figure*}

\vspace{-0.1in}\noindent \textbf{Case 1:} $B_{\mathit{new}}$ and
$B_{\mathit{curr}}$ are the same tuple block. Two tuple block $B_k$ and $B_l$
are considered to be the same if they cover the same set of categorical
patterns, e.g.\ $\forall X_i \in \mathcal{X}, \boldsymbol{x}_{i,j} \in
\mathcal{S}_{X_i}~s.t.~\tilde{p}(T = \boldsymbol{x}_{i,j}) \in \tilde{P}$, we
have $I_{X_i}(B_k \mid \boldsymbol{x}_{i,j}) = I_{X_i}(B_l \mid
\boldsymbol{x}_{i,j})$. Since block $B_{\mathit{new}}$ and $B_{\mathit{curr}}$
are the same and $B_{\mathit{curr}}$ is already part of the block graph,
inserting $B_{\mathit{new}}$ into block graph is not necessary any more. Thus,
we simply return \textit{Success} in this scenario (line~\ref{find:line:26}
\textendash~\ref{find:line:27}).\\ \vspace{-0.1in}

\noindent \textbf{Case 2:} $B_{\mathit{new}} \subseteq B_{\mathit{curr}}$. In
this case, the new tuple block $B_{\mathit{new}}$ should be inserted
between block $B_{\mathit{last}}$ and $B_{\mathit{curr}}$. To achieve this,
block $B_{\mathit{curr}}$ is first removed from the child block set of
$B_{\mathit{last}}$, and added as the child block of $B_{\mathit{new}}$.
Finally, the new block $B_{\mathit{new}}$ is inserted as the child block of
$B_{\mathit{last}}$, and \emph{Success} is returned (line~\ref{find:line:21}
\textendash~\ref{find:line:25}).\\ \vspace{-0.1in}

\noindent \textbf{Case 3:} $B_{\mathit{curr}} \subseteq B_{\mathit{new}}$. In
this scenario, the new tuple block $B_{\mathit{new}}$ should be inserted
as a descendant of the current block $B_{\mathit{curr}}$. Depending on whether
the block $B_{\mathit{curr}}$ has any child blocks, the insertion operation can
be further divided into two sub-cases:
\begin{itemize}[itemsep=-2pt, topsep=-2pt]
	\item \textbf{Case 3.1:} block $B_{\mathit{curr}}$ has no child block. In
		this scenario, the new block $B_{\mathit{new}}$ is directly inserted as
		the new child of $B_{\mathit{curr}}$ (line~\ref{find:line:2}
		\textendash~\ref{find:line:4});
	\item \textbf{Case 3.2:} block $B_{\mathit{curr}}$ has child blocks. Then,
		for each child block of $B_{\mathit{curr}}$, the \textit{findPosition}
		procedure is recursively performed to find the correct position to
		insert block $B_{\mathit{new}}$ (line~\ref{find:line:6}
		\textendash~\ref{find:line:14}). If none of these operations succeeds,
		block $B_{\mathit{new}}$ will be inserted as a new child block of
		$B_{\mathit{curr}}$ (line~\ref{find:line:15}
		\textendash~\ref{find:line:17}). At last, the descendants of the
		child blocks of $B_{\mathit{curr}}$ on which the \emph{findPosition}
		procedure failed to insert the block $B_{\mathit{new}}$ are further
		examined to see whether any of them would satisfy the partial order with
		block $B_{\mathit{new}}$ and be added as the child block of
		$B_{\mathit{new}}$ (line~\ref{find:line:18}).
\end{itemize}

\noindent \textbf{Case 4:} $B_{\mathit{new}}$ does not have any particular
relationship with $B_{\mathit{curr}}$. In this case, nothing needs to done with
the tuple blocks $B_{\mathit{curr}}$ and $B_{\mathit{new}}$, and \textit{Fail}
is simply returned to indicate that the attempt to insert block
$B_{\mathit{new}}$ is failed.

Figure~\ref{fig:4.A.1} shows an example of constructing such hierarchical block
graph on a small toy dataset with 6 attributes and 3 categorical patterns. With
the block graph $G$, the size of the tuple block could be easily calculated
using the set inclusion-exclusion principle. We first define the cumulative size
of a tuple block $B$, which is given by
$$
\mathit{cum}(B) = \prod_{A_i \in \mathcal{A} \setminus \attr(B)}
|\mathcal{R}(A_i)|~.
$$
Then the actual block size for block $B$ could be computed as
$$
|B| = \mathit{cum}(B) - \sum_{B_k \in \mathcal{B}_{\mathcal{X}}, B \subseteq
B_k} |B_k|~.
$$
In the block graph $G$, the tuple blocks that satisfy $B_k \in
\mathcal{B}_{\mathcal{X}}, B \subseteq B_k$ are simply those descendant blocks
of $B$. Algorithm~\ref{alg:4.A.4} describes the procedure of computing block
size for each tuple block in $\mathcal{B}_{\mathcal{X}}$ with the block graph
$G$, where $\mathit{desc}(B)$ represents the set of descendant blocks of $B$ in
the graph $G$.

\begin{algorithm}[t]
	\SetAlgoLined
	\SetKwInOut{Input}{input}
	\SetKwInOut{Output}{output}
	\SetKwFunction{KwFnCompute}{computeBlockSize}

	\Input{tuple block graph $G$, current visited block
		$B_{\mathit{curr}}$.}
	\Output{Block size for each $B \in \mathcal{B}_{\mathcal{X}}$.}
	\BlankLine

	$\mathit{cum}(B_{\mathit{curr}}) \leftarrow \prod\limits_{A_i \in
	\mathcal{A} \setminus \attr(B_{\mathit{curr}})} |\mathcal{R}(A_i)|$\;
	\If{$\mathit{child}(B_{\mathit{curr}}) = \emptyset$}{
		$|B_{\mathit{curr}}| \leftarrow \mathit{cum}(B_{\mathit{curr}})$\;
		\Return\;
	}
	\ForEach{$B_k \in \mathit{child}(B_{\mathit{curr}})$}{
		\KwFnCompute{$G$, $B_k$}\;
	}
	$|B_{\mathit{curr}}| \leftarrow \mathit{cum}(B_{\mathit{curr}}) -
	\sum\limits_{B_k \in \mathit{desc}(B_{\mathit{curr}})} |B_k|$\;
	\Return\;

	\caption{\textit{computeBlockSize} procedure}
	\label{alg:4.A.4}
\end{algorithm}

When individual attribute constraints are taken into account, the problem become
a little more complicated. However, it is obviously not feasible to combine the
individual attribute constraints with the categorical pattern constraints
together and construct the tuple block graph. This will make the tuple block
space blow up. Instead, as we mentioned previously in
Section~\ref{sec:column_margin}, the individual attribute constraints are
modeled with a separate Maximum Entropy distribution $p_{\mathcal{A}}$,
defined in Equation~\eqref{eq:3.B.2}, which only considers these constraints.
The block graph $G$ is still constructed based on the categorical patterns in
$\mathcal{X}$, which will exactly have the same structure as before. In this
case, following the same logic, the probability for tuple block $B$
becomes
\begin{align*}
	p(B) = p_{\mathcal{A}}(B) \cdot \frac{u_0}{v_0} \cdot \prod_{X_i \in
		\mathcal{X}} \prod_{\boldsymbol{x}_{i,j} \in \mathcal{S}_{X_i}}
		(u_{i,j})^{I_{X_i}(B \mid \boldsymbol{x}_{i,j})}~,
\end{align*}
where $p_{\mathcal{A}}(B) = \sum_{T \in B} p_{\mathcal{A}}(T)$ denotes the
probability of tuple block $B$ under the separate Maximum Entropy
distribution $p_{\mathcal{A}}$. Thus, the problem of computing the probability
$p(T = \boldsymbol{x}_{i,j})$ in becomes calculating probabilities of tuple
blocks $p_{\mathcal{A}}(B)$ for each $B \in \mathcal{B}_{\mathcal{X}}$. Since
$p_{\mathcal{A}}$ only takes the individual attribute constraints into account,
every attribute is independent of each other under the Maximum Entropy
distribution $p_{\mathcal{A}}$. Similar to the cumulative size of a tuple block,
we define the cumulative probability of a tuple block under $p_{\mathcal{A}}$ as
$$
p_{\mathcal{A}}^{(c)}(B) = \prod_{A_i \in \attr(B)} p_{\mathcal{A}}\left(T =
a^{(i)}_{j}\right)~,
$$
where $a_j^{(i)}$ is the value of attribute $A_i$ associated with tuple
block $B$. With the exponential form described in Equation~\eqref{eq:3.B.2}, it
is not difficult to verify that the probability of $T = a_{j}^{(i)}$ under
Maximum Entropy distribution $p_{\mathcal{A}}$ is:
$$
p_{\mathcal{A}}\left(T = a_j^{(i)}\right) = \frac{v_{i,j}}{\sum_{l = 1}^{k_i}
v_{i,l}}~.
$$
Again, to compute $p_{\mathcal{A}}(B)$ for all $B \in \mathcal{B}_{\mathcal{X}}$
with the set inclusion-exclusion principle, we could directly apply the
\textit{computeBlockSize} procedure with $|B|$ and $\mathit{cum}(B)$ replaced
by $p_{\mathcal{A}}(B)$ and $p_{\mathcal{A}}^{(c)}(B)$ respectively.

Notice that the model parameters $v_{i,j}$ also need to be updated in the
Iterative Scaling framework. However, the block graph $G$ is constructed without
considering individual attribute patterns, which makes it difficult to compute
the probabilities of these individual attribute patterns under the Maximum
Entropy model directly from the block graph $G$. In order to get these
probabilities, we treat these individual attribute patterns as arbitrary
categorical patterns and query their probabilities from the Maximum Entropy
model. The detail of querying the Maximum Entropy model will be described in
the following section.

Finally, the model inference algorithm could be further optimized in the
following way. Suppose the categorical patterns in $\mathcal{X}$ could be
divided into two disjoint groups, e.g.\ $\mathcal{X}_1, \mathcal{X}_2 \subset
\mathcal{X}$ and $\mathcal{X}_1 \cup \mathcal{X}_2 = \mathcal{X}$ such that
$\forall X_1 \in \mathcal{X}_1, \forall X_2 \in \mathcal{X}_2$ we have $X_1 \cap
X_2 = \emptyset$. In this case, the Maximum Entropy model $p^{*}_{\mathcal{X}}$
over $\mathcal{X}$ could be factorized into two independent components
$p^{*}_{\mathcal{X}_1}$ and $p^{*}_{\mathcal{X}_2}$ such that
$p^{*}_{\mathcal{X}} = p^{*}_{\mathcal{X}_1} \cdot p^{*}_{\mathcal{X}_2}$.
Furthermore, $p^{*}_{\mathcal{X}_1}$ and $p^{*}_{\mathcal{X}_2}$ only rely on
pattern set $\mathcal{X}_1$ and $\mathcal{X}_2$, respectively. Such
decomposition greatly reduces the sizes of tuple block spaces
$\mathcal{B}_{\mathcal{X}_1}$ and $\mathcal{B}_{\mathcal{X}_2}$ compared to the
original $\mathcal{B}_{\mathcal{X}}$, and could also be extended to the scenario
when there are multiple such disjoint pattern groups. Due to the independence
between these Maximum Entropy components, they can also be inferred parallelly
to further speed up the model inference process.

\subsection{Querying the Model}
\label{sec:query}
Given an arbitrary categorical pattern $X^{\prime} \notin \mathcal{X}$ with
associated value $\boldsymbol{x}^{\prime}$, to query the probability under the
Maximum Entropy distribution $p^{*}$, we perform the following operations. Let
$\mathcal{X}^{\prime} = \mathcal{X} \cup \{X^{\prime}\}$, and a temporary
tuple block graph $G^{\prime}$ is constructed by applying the procedure
described in Algorithm~\ref{alg:4.A.2} over categorical pattern set
$\mathcal{X}^{\prime}$.  Then the size of each tuple block in graph
$G^{\prime}$ is computed by calling \textit{computeBlockSize} procedure, and the
probability of categorical pattern $X^{\prime}$ is given by
$$
p^{*}(T = \boldsymbol{x}^{\prime}) = \sum_{\substack{B \in
	\mathcal{B}_{\mathcal{X}^{\prime}} \\ I_{X^{\prime}}(B \mid
	\boldsymbol{x}^{\prime}) = 1}} p^{*}(B)~.
$$

\section{Model Selection}
\label{sec:selection}

\begin{algorithm}[t]
	\SetAlgoLined
	\SetKwInOut{Input}{input}
	\SetKwInOut{Output}{output}
	\SetKwFunction{KwFnIS}{Iterative\_Scaling}

	\Input{A set of categorical patterns $\mathcal{X}$, and associated empirical
	probabilities $\tilde{P}$.}
	\Output{A set of most informative patterns $\mathcal{X}^{\prime}$.}
	\BlankLine

	$\mathcal{X}^{\prime} \leftarrow \varnothing$\;
	$p^{*} \leftarrow$ \KwFnIS{$\mathcal{X}^{\prime}$}\;
	\While{$\mathit{BIC}_{\mathcal{X}^{\prime}}$ decreases}{
		$X^{\prime} \leftarrow \argmax\limits_{X \in \mathcal{X}} h\big(p^{*}(T
		= \boldsymbol{x}), \tilde{p}(T = \boldsymbol{x} \mid D)\big)$\;
		$\mathcal{X}^{\prime} \leftarrow \mathcal{X}^{\prime} \cup
		\{X^{\prime}\}$\;
		$p^{*} \leftarrow \KwFnIS{$\mathcal{X}^{\prime}$}$\;
	}
	\Return{$\mathcal{X}^{\prime}$}\;

	\caption{Heuristic search procedure for most informative prior patterns}
	\label{alg:5.1}
\end{algorithm}

In order to discover the most informative prior information from pattern set
$\mathcal{X}$, we adopt the
Bayesian Information Criterion (BIC),
defined as:
$$
\mathit{BIC}_{\mathcal{X}} = -2 \log \mathcal{L}_{\mathcal{X}} 
+ N \cdot \log |D|~,
$$
where $\mathcal{L}_{\mathcal{X}}$ denotes the log-likelihood of the Maximum
Entropy model inferred over pattern set $\mathcal{X}$, $N$ represents the number
of model parameters, and $|D|$ is the number categorical tuples in the dataset
$D$. With the exponential form of the Maximum Entropy distribution specified in
Equation~\eqref{eq:3.A.1}, its log-likelihood given dataset $D$ is equal to
\begin{align*}
	\mathcal{L}_{\mathcal{X}} = \sum_{T \in D} \log p^{*}(T)
	= |D| \bigg(\log u_0 + \sum_{X_i \in \mathcal{X}}
	\sum_{\boldsymbol{x}_{i,j} \in \mathcal{S}_{X_i}} \tilde{p}(T =
	\boldsymbol{x}_{i,j} \mid D) \cdot \log u_{i,j} \bigg)~.
\end{align*}


The ideal approach to select the most informative categorical patterns from
pattern set $\mathcal{X}$ would be finding a subset of $\mathcal{X}$ that
minimizes the BIC score of the model. However, notice that this approach
involves a number of model inference operations which is proportional to the
number of subsets of $\mathcal{X}$. Considering the computation required for the
model inference, this method may be infeasible in practice.  Hence, we resort to
heuristics. Basically, what we desire are the patterns whose empirical
frequencies diverge most from their probabilities under current Maximum Entropy
model. In this case, they will contain the most new information compared to what
the model already knows. Thus, we borrow idea from Kullback-Leibler (KL)
divergence, where we make the probability of the categorical pattern $X$ under
consideration as one term and the rest of the probability mass as the other
term. To be more specific, the heuristic we use is defined as
$$
h(\alpha, \beta) = \alpha \log \frac{\alpha}{\beta} + (1 - \alpha) \log \frac{1
- \alpha}{1 - \beta}~.
$$
Instead of directly searching in the space of power set of $\mathcal{X}$, we
adopt an iterative search strategy. Starting from the empty model without any
prior information, in each iteration, we choose the pattern $X \in \mathcal{X}$
that maximizes the heuristic $h(p^{*}(T = \boldsymbol{x}), \tilde{p}(T =
\boldsymbol{x} \mid D))$ to update the current Maximum Entropy model. Here,
$p^{*}(T = \boldsymbol{x})$ and $\tilde{p}(T = \boldsymbol{x} \mid D)$ denote
the probability of pattern $X$ under current Maximum Entropy model and its
empirical frequency in the given dataset $D$, respectively. As the model
incorporates more and more patterns in $\mathcal{X}$, it becomes more certain
about the data, and the negative log-likelihood decreases. However, the model
becomes more complicated at the same time, and the penalty term in BIC becomes
large. This procedure continues until the BIC score of the model does not
decrease any more. Algorithm~\ref{alg:5.1} describes the details of this
heuristic search approach.

\section{Experimental Results}
\label{sec:exp}

\begin{figure}[t]
	\centering\vspace{-0.5cm}
	\includegraphics[width=3in]{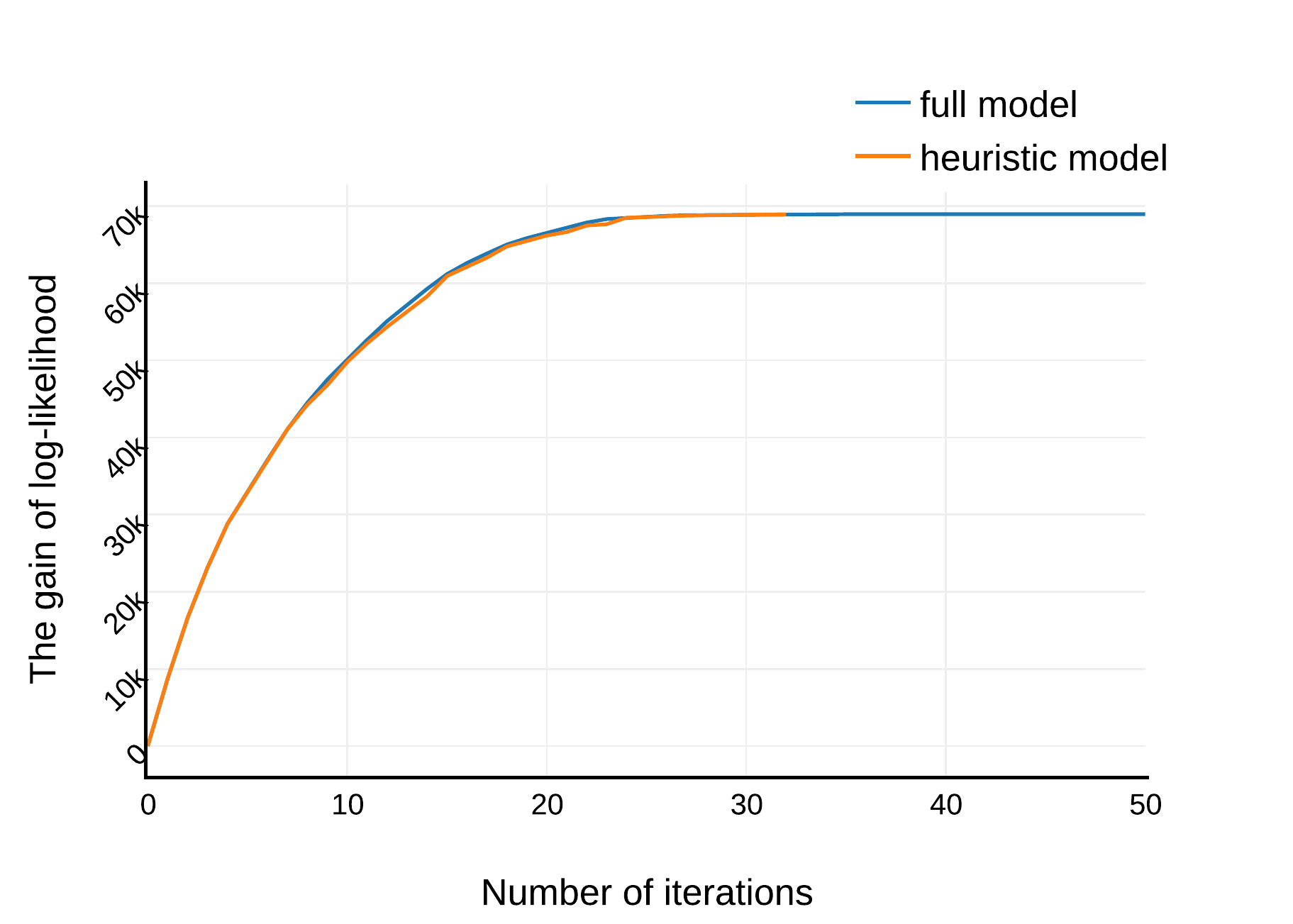}
	\caption{The gain of the log-likelihood of the full model and heuristic
		model compared to the based line model. The blue line and orange line
		are so close that they overlap with each other in some iterations. Also
		notice that orange line for heuristic model stop early due to the model
	selection with BIC.} \vspace{-0.4cm}
	\label{fig:6.A.1}
\end{figure}

\subsection{Synthetic Data Generation}
\label{sec:data_gen}
To evaluate the proposed Maximum Entropy model against the true generating
distribution of categorical data, we generate synthetic datasets. Usually when
the entire categorical data space is large, it is infeasible to specify an exact
generating distribution for categorical data. Thus, we generate the synthetic
data $D$ with the following approach.

A set of categorical attributes $\mathcal{A}$ is first generated, and the
number of possible values for each attribute $A_i \in \mathcal{A}$ is randomly
sampled from a given range. Each categorical attribute $A_i$ is associated with
a random generated probability distribution (marginal distribution) that
specifies the probability of each possible value of $A_i$. In order the enforce
the dependency between attributes, a set of categorical patterns $\mathcal{X}$
is generated and each of these pattern is associated with a probability. To
generate a categorical tuple in the synthetic dataset, we sample from a
Bernoulli distribution parameterized by the pattern frequency of each $X \in
\mathcal{X}$ to determine whether this tuple should contain this pattern or not.
For the rest of the attributes that are not covered by any of these patterns in
$\mathcal{X}$, their values in the generated categorical tuple are sampled
independently from their corresponding marginal distributions respectively. Such
process is repeated to obtain the desired number of categorical tuples in the
synthetic dataset. In our experiments, we set $|\mathcal{A}| = 100$, number of
patterns $|\mathcal{X}| = 50$, and the number categorical tuples in synthetic
dataset $|D| = 10,000$. All the experiments were conducted on a 80-core Xeon 2.4
GHz machine with 1 TB memory, and the results were averaged across 10
independent runs.

\begin{figure*}[t]
	\vspace{-0.5cm}
	\begin{minipage}[t]{0.48\textwidth}
		\centering
		\includegraphics[width=3in]{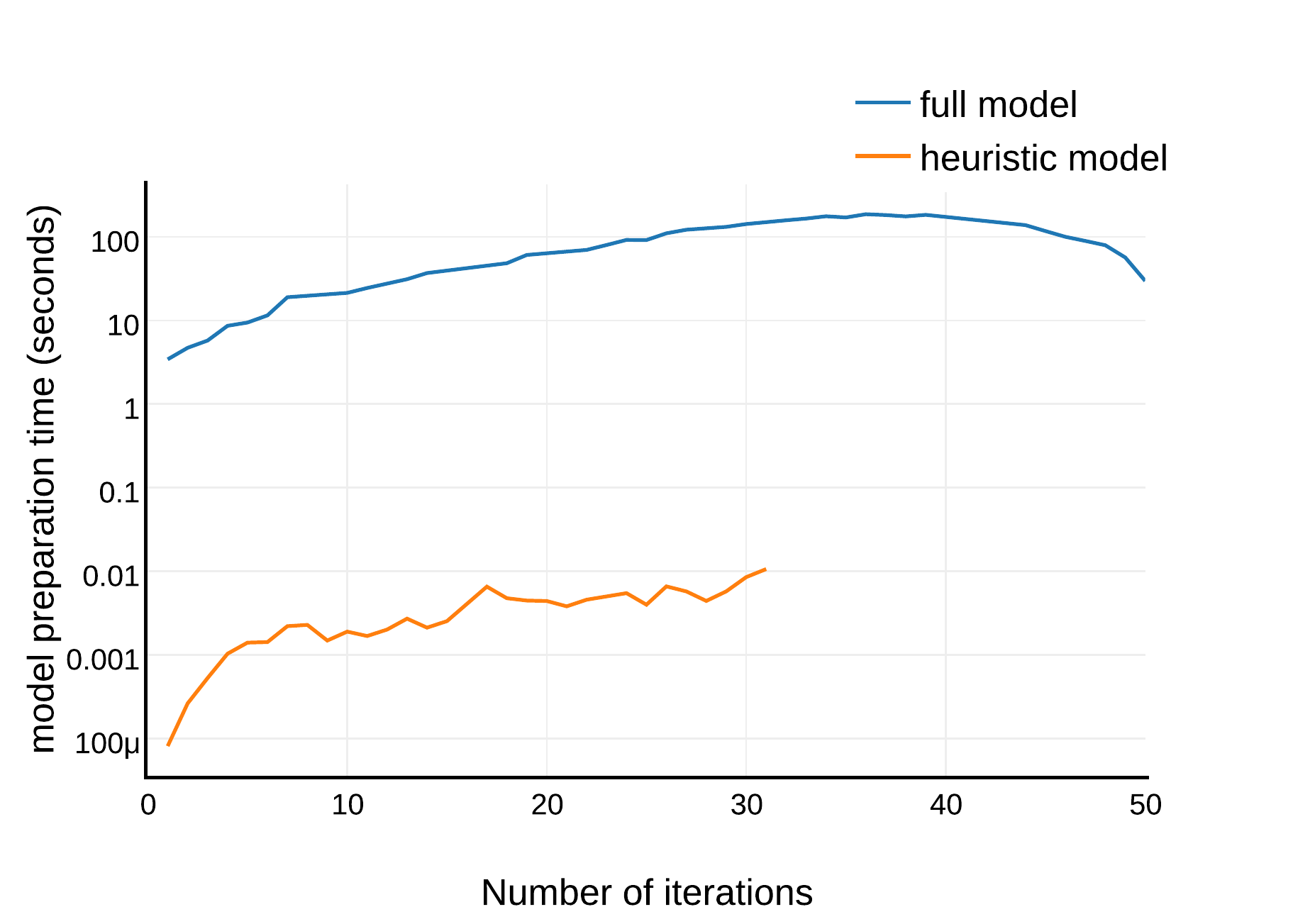}
		\caption{Model preparation time of each iteration as we iteratively
		choose the most informative patterns. Y-axis is in log scale.}
		\label{fig:6.A.3}
	\end{minipage}
	\hfill
	\begin{minipage}[t]{0.48\textwidth}
		\centering
		\includegraphics[width=3in]{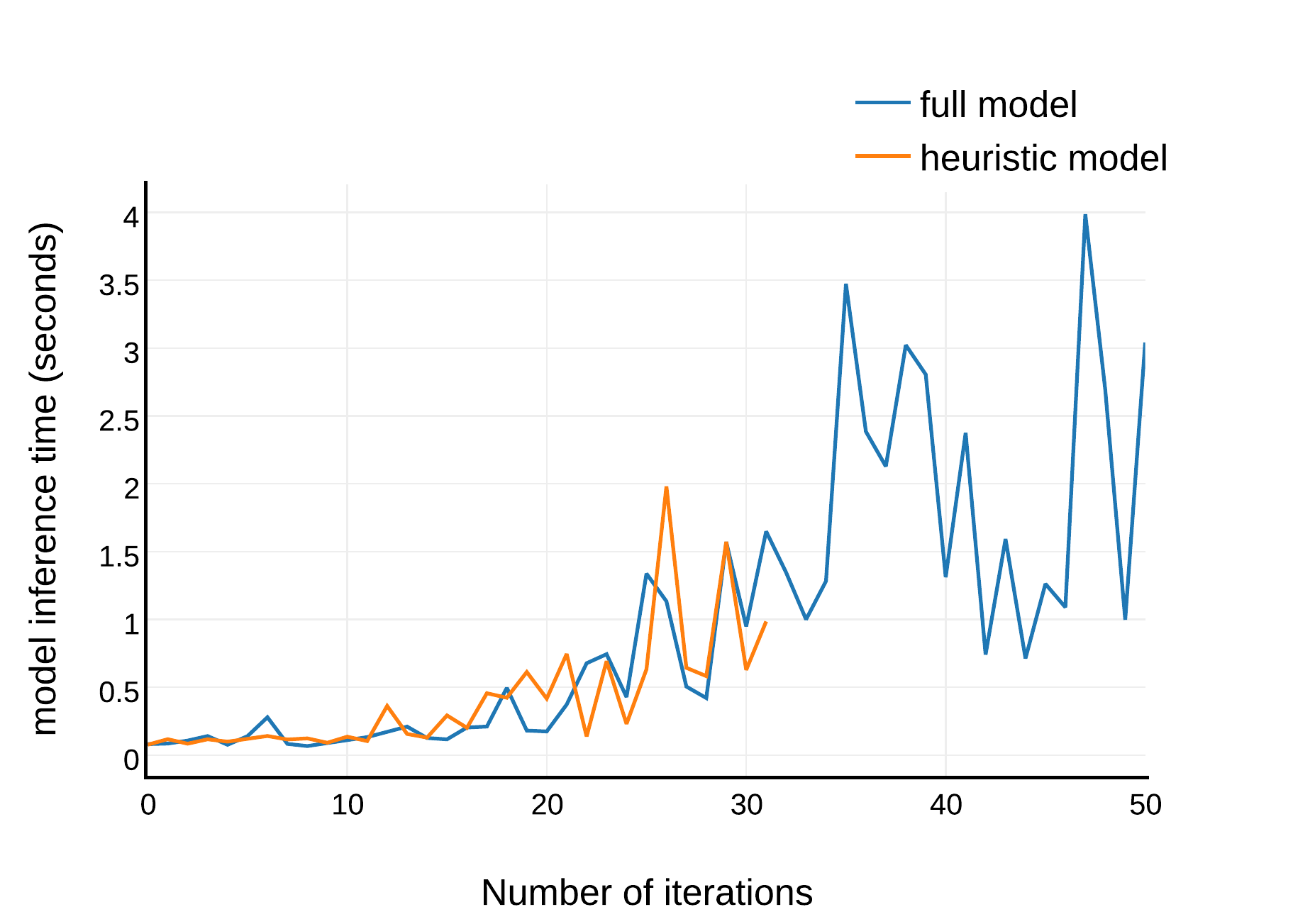}
		\caption{Model inference time of each iteration as we iteratively choose
		the most informative patterns.}
		\label{fig:6.A.2}
	\end{minipage}
	\vspace{-0.4cm}
\end{figure*}

\subsection{Results on Synthetic Data}
\label{sec:synthetic}

\begin{table}[!t]
	\centering
	\caption{Comparison of approximate KL-divergence measures between full
	model, heuristic model and baseline model.} 
	\label{tab:kl_compare}
	\begin{tabular}{c c c c}
		\toprule
		& full model & heuristic & baseline \\
		\midrule
		$\mathit{\hat{KL}}(p^{*}, p^{\prime})$ & $\mathbf{9.410 \times 10^{-5}}$
		& $8.566 \times 10^{-4}$ & 1.8881 \\
		$\mathit{\hat{KL}}(\tilde{p}~, p^{\prime})$ & \textbf{0.1695} & 0.1836
		& 2.0664 \\ 
		\bottomrule
	\end{tabular}
\end{table}

We first verify that the heuristic function $h(\alpha,\beta)$ proposed in
Section~\ref{sec:selection} could discover the most informative patterns from
$\mathcal{X}$ based on the current knowledge the model already knows. We refer
the Maximum Entropy model inferred with entire pattern set $\mathcal{X}$ and all
the individual attribute frequencies as \emph{full model}, and the Maximum
Entropy model selected by heuristic and BIC as \emph{heuristic model}. Notice
that in the heuristic model, individual attribute frequencies are also taken
into account. In this experiment, we iteratively updated the model with the
patterns in $\mathcal{X}$, and measured the log-likelihood in each iteration.
However, using BIC to select the model may result different number of patterns
incorporated over different synthetic datasets. Thus, we report the results over a
single synthetic dataset here. For the full model, the pattern in $\mathcal{X}$
that maximized the log-likelihood in each iteration were selected and added to the
model. 
\begin{table}[!t]
	\centering
	\caption{Comparison of model preparation time ($t_{\mathit{pre}}$), model
	inference time ($t_{\mathit{infer}}$) and data sampling time
	($t_{\mathit{sample}}$) between full model and heuristic model (in seconds).}
	\label{tab:time_compare}
	\begin{tabular}{c c c c}
		\toprule
		& $t_{\mathit{pre}}$ & $t_{\mathit{infer}}$ & $t_{\mathit{sample}}$
		\\ \midrule
		full model  & 4438.785 & 27.266 & 1.678 \\
		heuristic model & \textbf{17.981} & \textbf{8.950} & \textbf{0.461}
		\\ \bottomrule
	\end{tabular}\vspace{-0.2in}
\end{table}


Figure~\ref{fig:6.A.1} illustrates the gain of the log-likelihood as the
model incorporates more and more patterns in $\mathcal{X}$. As expected, the
gain of the log-likelihood of the full model is larger in some iterations since
it identifies the optimal pattern in each iteration with respect to the
likelihood.  We also observe that although not optimal, the log-likelihood of
the heuristic model approximates that of the full model quite well, which
demonstrates that the proposed heuristic successfully identifies the relatively
informative patterns in each iteration. In the last few iterations, the gain
of log-likelihood of the full model barely changes. This indicates that the
patterns selected in these iterations are less informative or even redundant.

To assess the quality of the reconstruction, we aim to apply KL divergence measures.
However, in practice, it is very difficult to compute the KL divergence between
the entire Maximum Entropy distribution and data generating distribution for the
categorical data due to the large categorical tuple space. As a trade off, we
use the probabilities of patterns in pattern set $\mathcal{Y}$ to characterize
the probability distributions for categorical data in both scenarios, and define
the following approximate KL-divergence measure:
\begin{align*}
	\mathit{\hat{KL}}(p^{*}, p^{\prime}) = \sum_{X \in \mathcal{Y}} \bigg[ 
		p^{*}(X) \log \frac{p^{*}(X)}{p^{\prime}(X)} + (1 - p^{*}(X)) \log
		\frac{1 - p^{*}(X)}{1 - p^{\prime}(X)}\bigg]~.
\end{align*}
Here, $p^{*}$ and $p^{\prime}$ denote the Maximum Entropy distribution and data
generating distribution respectively, and pattern set $\mathcal{Y}$ could be
only categorical pattern set $\mathcal{X}$ or $\mathcal{X} \cup \mathcal{A}$ if
individual attribute frequencies are considered. We also compute the
$\mathit{\hat{KL}}(\tilde{p}, p^{\prime})$ to compare the empirical probability
distribution, say $\tilde{p}$, in the samples generated by the categorical
Maximum Entropy model with the true data generating distribution. In this
experiment, we computed $\mathit{\hat{KL}}(p^{*}, p^{\prime})$ and
$\mathit{\hat{KL}}(\tilde{p}, p^{\prime})$ for both full model and heuristic
model. For comparison purpose, we used independent attribute model
$p_{\mathcal{A}}$ where each categorical attribute is independent of each other
as the baseline model. For each of these models under consideration, 1000
categorical data samples were generated to compute empirical probability
distribution $\tilde{p}$.

\begin{figure*}[t]
	\vspace{-0.5cm}
	\begin{minipage}[t]{0.32\textwidth}
		\centering
		\includegraphics[width=2.3in]{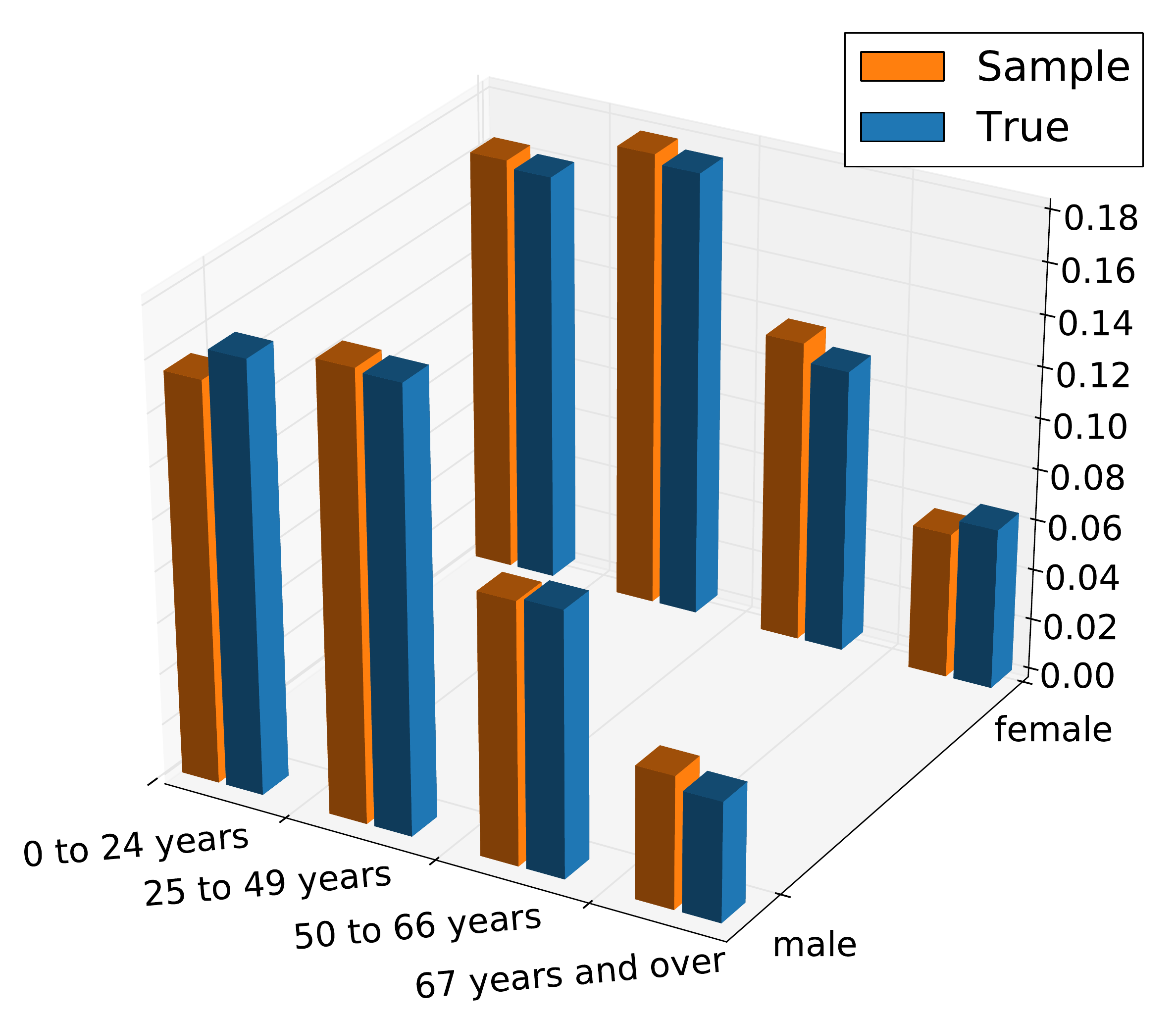}
	\end{minipage}
	\hfill
	\begin{minipage}[t]{0.32\textwidth}
		\centering
		\includegraphics[width=2.3in]{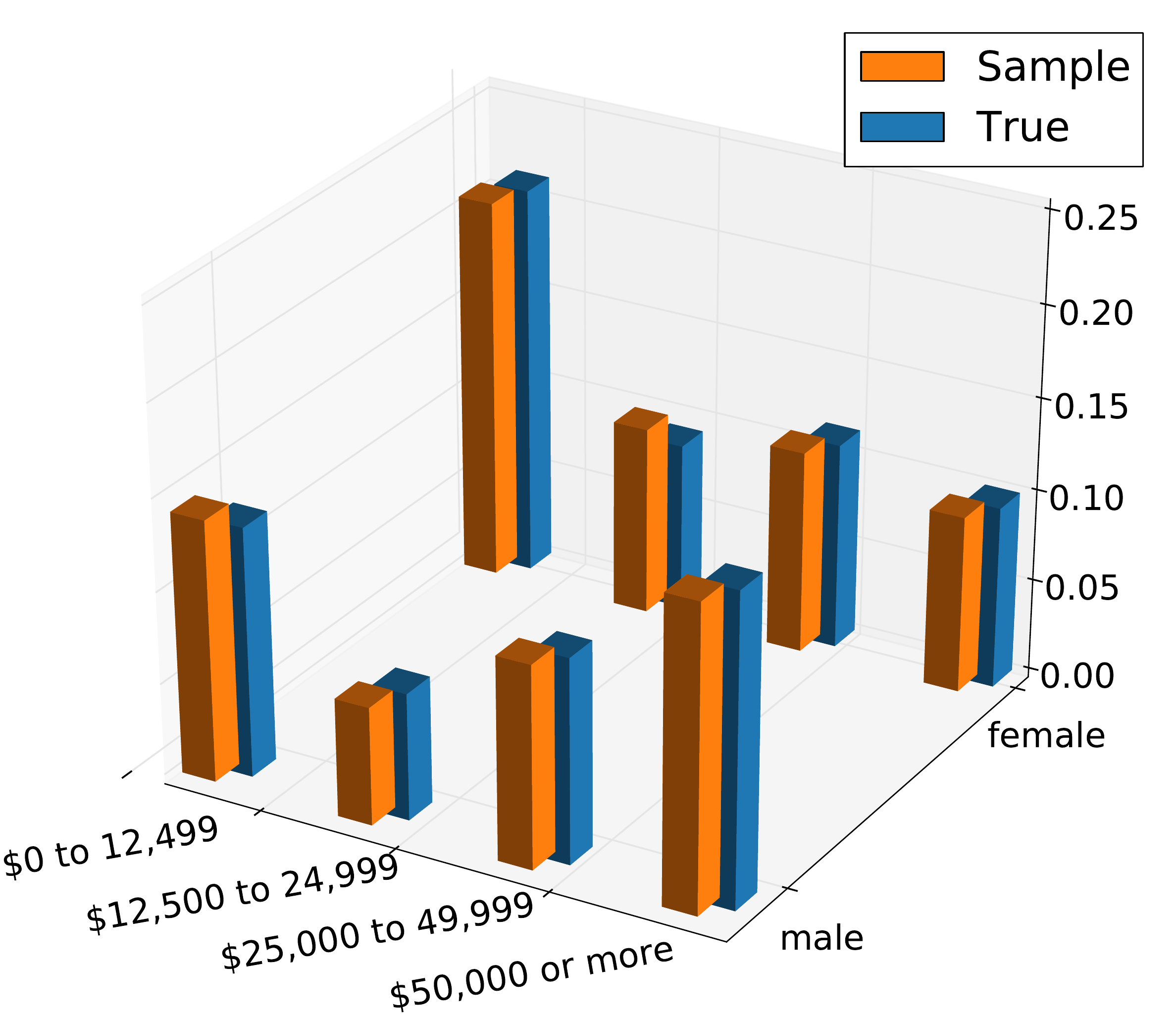}
	\end{minipage}
	\hfill
	\begin{minipage}[t]{0.32\textwidth}
		\centering
		\includegraphics[width=2.3in]{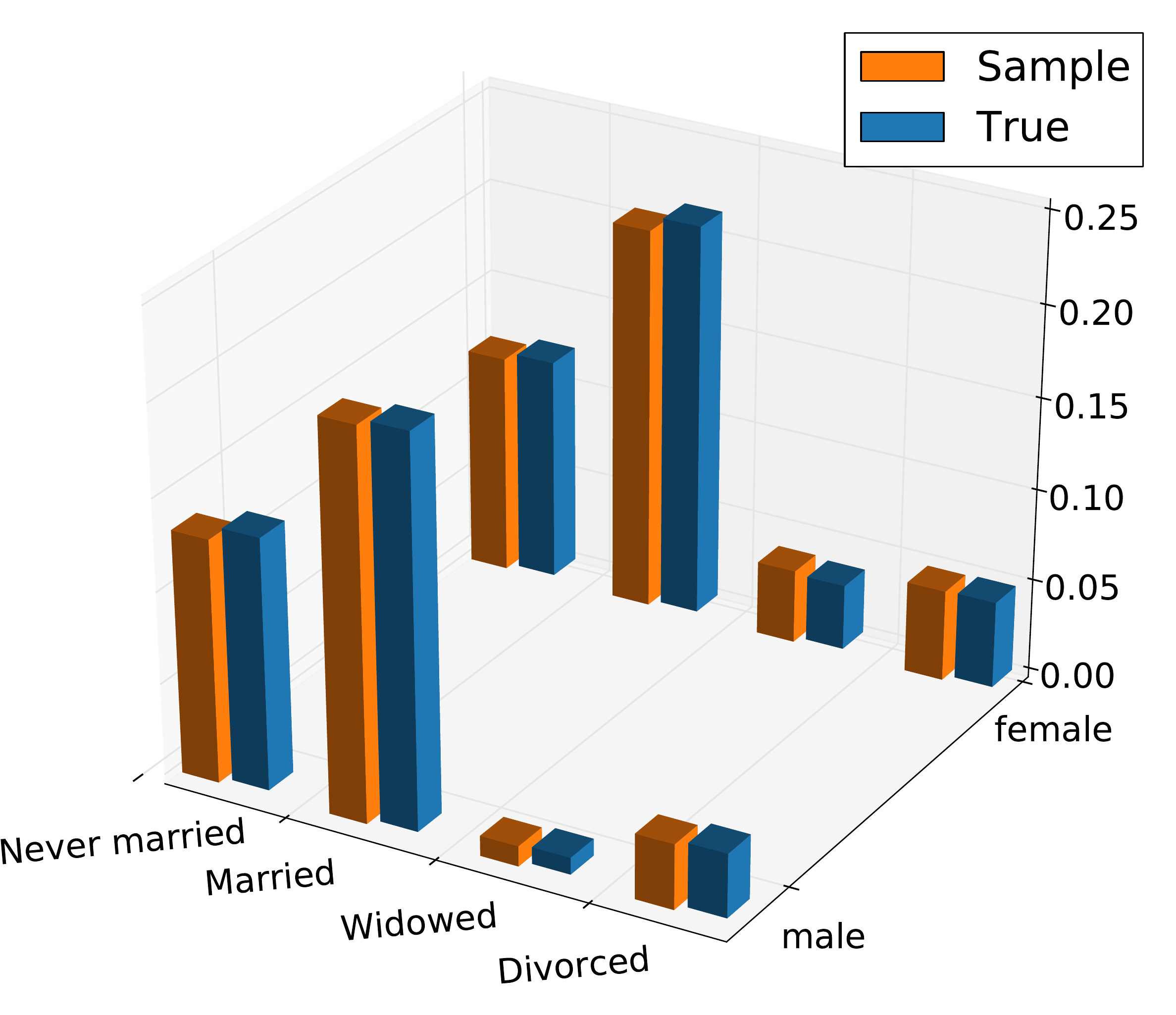}
	\end{minipage}
	\caption{Comparison of two-attribute marginal distributions between true
		statistics in Virginia ACS summary data and samples generated by
		categorical Maximum Entropy model for categorical patterns \{\emph{sex,
		age}\} (left), \{\emph{sex, income}\} (middle), and \{\emph{sex, marital
		status}\} (right). For pattern \{\emph{sex, marital status}\}, the
		pattern values whose \emph{marital status} is \emph{Others under 15
		years old} is not displayed here since for those individuals, their
		marital statuses are unavailable.} \vspace{-0.5cm}
	\label{fig:6.C.4}
\end{figure*}
Table~\ref{tab:kl_compare} compares these approximate KL-divergence measures.
In Table~\ref{tab:kl_compare}, the small approximate KL-divergence values for
full model and heuristic model indicate that the categorical Maximum Entropy
distributions converge to the underlying data generation distribution. Moreover,
the samples generated by these two models also successfully maintain the
properties of the data generation distribution. This demonstrates that our model
is capable of recovering the true categorical data distribution. When compared
to the baseline model, our model outperforms several magnitudes in term of
estimation accuracy.

We also measure the time required to prepare the pattern set that serves as
prior information of the model $t_{\mathit{pre}}$, the time to infer the Maximum
Entropy model $t_{\mathit{infer}}$, and the time to sample a single categorical
tuple from the model $t_{\mathit{sample}}$. Here, for the full model,
$t_{\mathit{pre}}$ refers to the time required to arrange the pattern set
$\mathcal{X}$ into the same order used in the iterative model update procedure
in the first experiment where the categorical pattern that maximizes the
log-likelihood is chosen in each iteration. Table~\ref{tab:time_compare}
compares the runtime performance between the full model and the heuristic model,
and Figure~\ref{fig:6.A.3} and~\ref{fig:6.A.2} show the $t_{\mathit{pre}}$ and
$t_{\mathit{infer}}$ of every iteration in the iterative procedure used to
verify the heuristic function $h(\alpha, \beta)$ in our first experiment. With
the informative as well as simple model selected by the heuristic function
$h(\alpha, \beta)$ and BIC, the heuristic model requires much less time to infer
the Maximum Entropy distribution and sample categorical tuples from the model.

\begin{figure}[!t]
	\centering\vspace{-0.7cm}
	\includegraphics[width=3in]{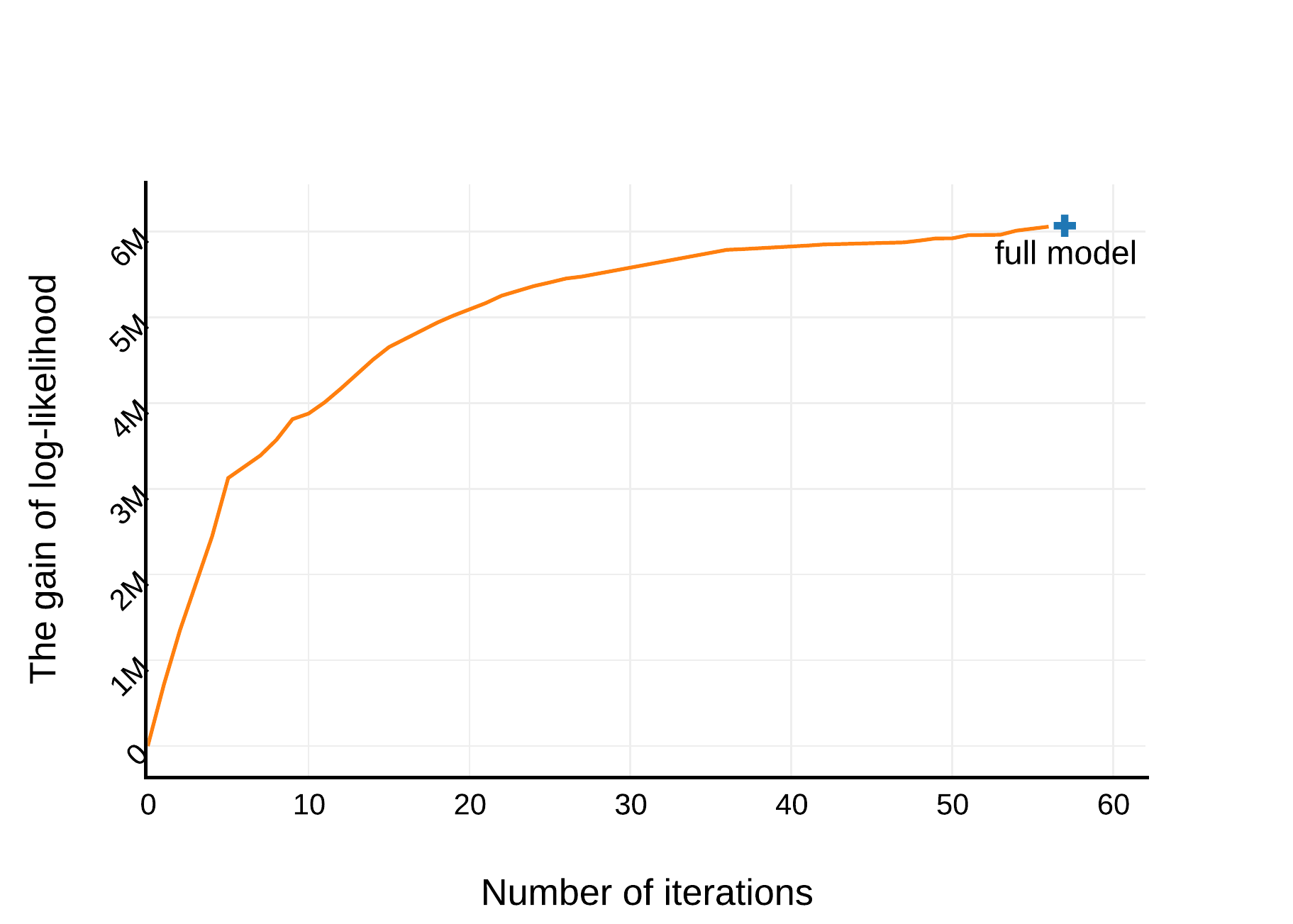}
	\caption{The gain of the negative log-likelihood of the model compared to
		the baseline model (model at iteration 0) over the Virginia ACS summary
		data. The data point marked with cross denotes the negative
		log-likelihood of the full model where all the categorical patterns in
	Virginia ACS summary data are considered.}\vspace{-0.3cm}
	\label{fig:6.C.1}
\end{figure}

\subsection{Results on Real Data}
\label{sec:real}

To evaluate the performance of the proposed categorical Maximum Entropy model
on the real data, we studied the problem of generating synthetic populations
with US census data. 
Specifically, we use the
2012 American Community Survey (ACS) 1-year summary data~\cite{us:census:acs}, which contains
aggregated statistics about age, sex, race,
income, and many other features. Some of these features, e.g.\ sex and race, are
perfect categorical attributes for the proposed Maximum Entropy model. While
although some other features, e.g.\ age and income, are numerical, they are
binned into several ranges based on their values, and treated here
as categorical attributes.

\begin{table}[t]
	\centering
	\captionof{table}{Top categorical patterns selected by heuristic
	model in Virginia ACS summary data.}
	\label{tab:ACS_patterns}
	\begin{tabular}{c c c}
		\toprule
		patterns & {number of possible values} &
		{number of selected values} \\
		\midrule
		\parbox{4cm}{\{means of transportation to work, occupation\}}
		& 49 & 34 \\
		\{sex, income\} & 8 & 2 \\
		\{sex, marital status\} & 10 & 2\\
		\{sex, age\} & 8 & 1\\
		\bottomrule
		\vspace{-0.3in}
	\end{tabular}
\end{table}

In our experiments, we chose the state of Virginia as our study case. Among all the
features in the ACS summary data, we selected \emph{sex}, \emph{age},
\emph{race}, \emph{income}, \emph{occupation}, \emph{marital status},
\emph{means of transportation to work}, \emph{education level}, and \emph{health
insurance coverage} as the set of categorical attributes. We converted the
corresponding aggregated statistics in ACS summary data into categorical
patterns, and inferred the heuristic model over these patterns.
Figure~\ref{fig:6.C.1} describes the gain of the log-likelihood of
the heuristic model, and the approximate KL-divergence measure between the
inferred Maximum Entropy distribution and the empirical data distribution in
Virginia ACS summary data is $0.0001975$. Table~\ref{tab:ACS_patterns} also
shows some most informative patterns selected by the proposed heuristic. Notice
that in Figure~\ref{fig:6.C.1}, the last data point marked with cross indicates
the gain of the log-likelihood of the full model where all the categorical
patterns in the Virginia ACS summary data are taken into account. 

We also generated a sample of $3,000$ synthetic individuals with the inferred
heuristic model for Virginia, and calculated the marginal distributions for all
individual attributes and some selected two-attribute categorical patterns.
Notice that for attributes \emph{Marital status}, \emph{Means of transportation
to work}, \emph{Occupation} and \emph{Education level}, the population
considered in the ACS summary data is not the entire population of Virginia
state. Thus, we add an additional value for these attributes, e.g.\ the value
\emph{Others under 15 years old} for the attribute \emph{Marital status}, to
denote the proportion of the entire population that are not taken into account
in the ACS summary data. 
Fig.~\ref{fig:6.C.4} show some of these marginal distributions and compares them
with Virginia ACS summary data. We can see that the empirical distributions
calculated from the synthetic individuals are very close to those in the
Virginia ACS summary data. Such results demonstrate that our categorical Maximum
Entropy model well maintains the statistical characteristics of the real world
datasets, and is capable of generating synthetic data for real applications.

\subsection{Application: Epidemic Simulation}
\label{sec:epi_sim}

\begin{figure}[!t]
	\centering\vspace{-0.23cm}
	\includegraphics[width=3in]{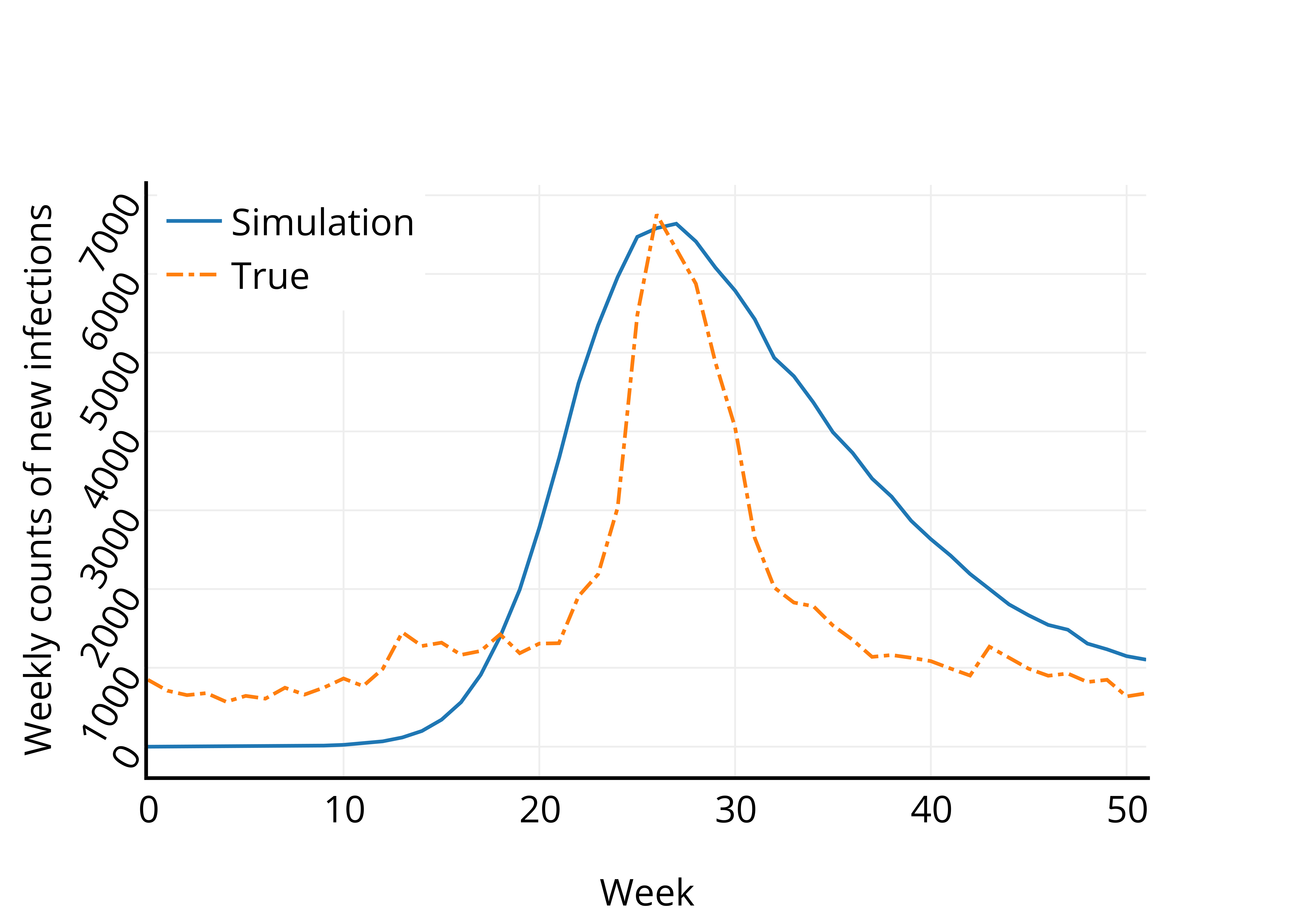}
	\caption{The simulated weekly flu new infection counts compared to the
		estimated weekly new infection counts from Google Flu Trends. The
	simulation results are averaged across 10 independent runs.}\vspace{-0.6cm}
	\label{fig:6.4.1}
\end{figure}

In this section, we apply our proposed categorical Maximum Entropy model to
generate synthetic population for the city of Portland OR in the United States,
and use this model for an epidemiological simulation.  We first take a publicly
available synthetic contact network dataset of Portland~\cite{NDSSL-TR-07-003},
which contains both individual demographic and contact information of the
residents in the city of Portland. The demographic information in this dataset
contains gender, age and household income. We first group the values of age and
household income into several ranges and change them into categorical features,
similar to our ACS dataset analysis in Section~\ref{sec:real}. Then we compute
the statistics, e.g.\ frequencies, of the single and pairwise demographic
features, convert them into categorical patterns, and infer the categorical
Maximum Entropy model over these patterns. The Portland dataset contains
$1,575,861$ connected individuals, where each individual performs at least one
activity with others. To generate our synthetic population for the Portland
dataset, we draw $1,575,861$ samples from the inferred categorical Maximum
Entropy model.

To construct the contact network for the synthetic population, we first match
the generated synthetic individuals to the real ones involved in the contact
activities described in the Portland dataset based on their demographical
feature values. Then the contact network can be naturally created by connecting
the synthetic individuals according to the contact activities they involves in.
In this application, we choose to study the flu season in the city of Portland
during the period from June 2013 to June 2014. We retrieve the estimated weekly
counts of flu new infections for the city of Portland from Google Flu
Trends~\cite{google:flu:trend}, and apply the Susceptible-Infectious (SI)
epidemic model over the contact network to fit the curve of weekly flu new
infection counts. Figure~\ref{fig:6.4.1} illustrates the fitted curve using the
SI epidemic model. As the figure shows, the simulation results of the SI model
over the synthetic population capture the trend and the peak of the weekly flu
new infections in the city of Portland. These results demonstrate that the
synthetic population generated by the categorical Maximum Entropy model is a
useful model of population-level activity in cities.

\section{Related Work}
\label{sec:related}

The problem of generating synthetic data that maintain the structures and
dependencies in actual data has been studied by the researchers from various
realms, ranging from network analysis to privacy
preservation.
The work in~\cite{Barrett:5429425} studied and analyzed large synthetic social
contact networks where the synthetic population was generated by applying 
iterative proportion fitting (IPF) techniques over census
data. Variants of IPF, e.g.\ hierarchical IPF~\cite{mueller2011hierarchical} and
two-stage IPF~\cite{zhu2014synthetic}, were also developed for generating synthetic
population data for various research purposes such as land use and
transportation microsimulation. Compared to the IPF-based
approach,~\citet{ma2015synthetic} proposed a fitness-based synthesis method to
directly generate synthetic population, and~\citet{barthelemy2013synthetic}
introduced a sample-free synthetic population generator by using the data at the
most disaggregated level to define the joint distribution. Besides generating
synthetic population with the combinational optimization based
technique,~\citet{namazi2014generating} also projected dynamics over the
synthetic population using a dynamic micro-simulation model. The Network
Dynamics and Simulation Science Laboratory at Virginia Tech released synthetic
datasets of population in the city of Portland~\cite{NDSSL-TR-07-003} and ad-hoc
vehicular radio network in Washington D.C.~\cite{NDSSL-TR-07-010}, which are
generated by the high-performance, agent-based simulation system
\textit{Simfrastructure}.
Recently,~\citet{Park:6680524} proposed a non-parametric data synthesizing
algorithm, particularly a perturbed Gibbs sampler, to generate large-scale
privacy-safe synthetic health data. Instead of using patterns to characterize
the data, a set of perturbed conditional probability distributions were
estimated to represent the data distribution.

In the database community, there exists several research work that generates
synthetic relational databases. For a
survey,~\citet{Gray:1994:QGB:191839.191886} discuss several database generation
techniques that generate large scale synthetic datasets, and
\citet{Bruno:2005:FDG:1083592.1083719} proposed a Data Generation Language (DGL)
that allows individual attribute distribution to be specified.
In~\cite{Houkjaer:2006:SRD:1182635.1164254}, the authors described a graph model
directed database generation tool which could handle complex inter- and
intra-table relationships in large database schemas.
~\citet{Arasu:2011:DGU:1989323.1989395} proposed an efficient, linear
programming based algorithm to generate synthetic relational databases that
satisfy a given set of declarative constraints.

There is also extensive work related to the topic of query optimization that
applies the Maximum Entropy principle in the database
community.~\cite{Kaushik:2009:GDS:1616173.1616179} and
~\cite{Markl:2005:CES:1083592.1083638} estimated the sizes of database
queries by modeling complicated database statistics using Maximum Entropy
probability distribution.~\citet{Re:2010:UCE:1807085.1807095} studied the
problem of cardinality estimation using the Entropy Maximization technique, and
proposed to use peak approximation to compute the approximate Maximum Entropy
distribution. In~\cite{Srivastava:2006:ICH:1129754.1129899}, the authors
described an algorithm called ISOMER which approximated the true data
distribution by applying the Maximum Entropy principle over the information
gained from database query feedback.

\section{conclusion}
\label{sec:conclusion}
In this paper, we propose a generative probabilistic model for the categorical
data by employing Maximum Entropy principle. By introducing categorical tuple
blocks and the corresponding partial order over them, we present an efficient
model inference algorithm based on the well-known Iterative Scaling framework.
Experiment results on both synthetic data and real US census data show that the
proposed model well estimates the underlying categorical data distributions. The
application to the problem of synthetic population generation demonstrates the
potential of the proposed model to help the researchers in various areas.

\section*{Acknowledgments}
Supported by the Intelligence Advanced Research Projects Activity (IARPA) via
DoI/NBC contract number D12PC000337, the US Government is authorized to
reproduce and distribute reprints of this work for Governmental purposes
notwithstanding any copyright annotation thereon. Disclaimer: The views and
conclusions contained herein are those of the authors and should not be
interpreted as necessarily representing the official policies or endorsements,
either expressed or implied, of IARPA, DoI/NBC, or the US Government.

\bibliographystyle{plainnat}
\bibliography{reference}

\end{document}